\pdfoutput=1

\documentclass[11pt]{article}

\usepackage{emnlp2021}

\usepackage{times}
\usepackage{latexsym}
\usepackage{hyperref}
\usepackage{url}
\usepackage{graphicx} 
\usepackage{amsmath}
\usepackage{multirow}
\usepackage{multicol}
\usepackage{color}
\usepackage{booktabs}

\usepackage[T1]{fontenc}

\usepackage[utf8]{inputenc}

\usepackage{microtype}

\title{Grid-VLP: Revisiting Grid Features for Vision-Language Pre-training }

\author{Ming Yan, Haiyang Xu, Chenliang Li, Bin Bi, Junfeng Tian, Min Gui and Wei Wang
 \\
 Alibaba Group 
\\
         ym119608,shuofeng.xhy,lcl193798,b.bi@alibaba-inc.com
\\
         tjf141457,guimin.gm,hebian.ww@alibaba-inc.com}

\begin{document}
\maketitle
\begin{abstract}
Existing approaches to vision-language pre-training (VLP) heavily rely on an object detector based on bounding boxes (regions), where salient objects are first detected from images and then a Transformer-based model is used for cross-modal fusion. Despite their superior performance, these approaches are bounded by the capability of the object detector in terms of both effectiveness and efficiency. Besides, the presence of object detection imposes unnecessary constraints on model designs and makes it difficult to support end-to-end training. 

In this paper, we revisit grid-based convolutional features for vision-language pre-training, skipping the expensive region-related steps. We propose a simple yet effective grid-based VLP method that works surprisingly well with the grid features. By pre-training only with in-domain datasets, the proposed Grid-VLP method can outperform most competitive region-based VLP methods on three examined vision-language understanding tasks. We hope that our findings help to further advance the state of the art of vision-language pre-training, and provide a new direction towards effective and efficient VLP.


\end{abstract}

\section{Introduction}
Recent studies~\citep{lu2019vilbert,su2019vl,tan2019lxmert,chenuniter,li2020oscar,yu2020ernie} on vision-language pre-training have substantially advanced the state of the art across a variety of vision-and-language (V+L) tasks. These approaches typically follow a two-stage pipeline: 1) a pre-trained object detector is first used to encode an image by identifying a set of salient visual objects from the image; and 2) a cross-modal fusion model is pre-trained to learn the cross-modal representations. While most recent VLP methods utilize region-based visual features~\citep{anderson2018bottom} extracted by object detection, this paper explores a new way for vision-language pre-training, by using vanilla grid-based convolutional features from ConvNets~\citep{he2016deep,xie2017aggregated}.


Despite their superior performance, region-based VLP solutions suffer from the three problems: 1) The task-specific object detector has a strong impact on the performance of the VLP models. Important visual information may be lost during object detection;
2) In object detection, the extraction and post-processing of region features is very time-consuming;
3) The feature extraction process is non-differentiable, which imposes unnecessary constraints on model designs and makes it difficult to support end-to-end training. 

To address the limitations, we revisit the grid-based convolutional features for vision-language pre-training. Specifically, we encode an image into a feature map with convolutional networks such as ResNet~\citep{he2016deep}, and then conduct cross-modal fusion directly between text and the derived image feature map. In this way, we step out of the bounding box design of local region features and make the full use of all visual information for vision and language learning. On the other hand, computing grid features enables more flexible and simpler model design and thus leads to inference speedup, by skipping expensive region-related steps.

In this paper, we propose a simple yet effective grid-based VLP method, namely Grid-VLP, by directly pre-training with the grid-based convolutional features instead of the dominant region-based features from bottom-up attention~\citep{anderson2018bottom}. We follow LXMERT~\citep{tan2019lxmert} and adopt the typical Masked Language Modeling, Image-Text Matching and Image Question Answering as our pre-training tasks. To handle the challenge of modeling long sequences of grid features and accelerate the training process, we utilize a random grid sampling mechanism as in PixelBERT~\citep{huang2020pixel}, which helps improve the robustness of visual feature learning. We pre-train Grid-VLP only with in-domain datasets (Visual Genome~\citep{krishna2017visual} and MS-COCO~\citep{lin2014microsoft}), and then fine-tune it on three popular V+L understanding tasks: VQA~\citep{antol2015vqa}, NLVR2~\citep{suhr2018corpus} and GQA~\citep{hudson2019gqa}. Our results show that with the grid features, Grid-VLP
outperforms the competitive region-based VLP methods across all tasks. Besides, we also provide in-depth analysis on the influence of different design choices regarding feature types, image encoder architectures and resolution of input images.


\begin{figure}
\centering
\includegraphics[width=0.49\textwidth]{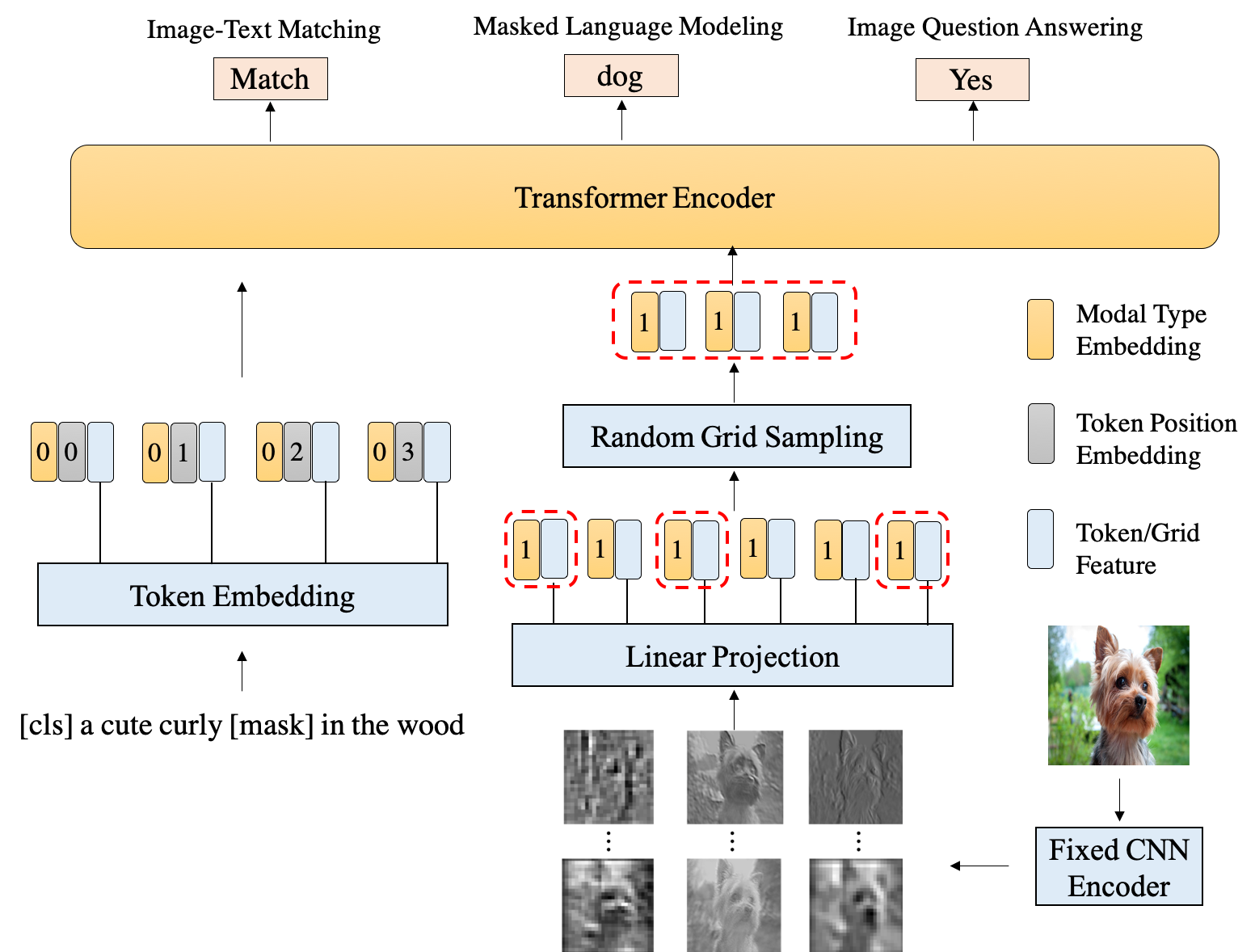}
\caption{The overall framework of Grid-VLP. } 
\label{fig:framework} 
\end{figure}

\section{Related Work}
Existing approaches to VLP~\citep{li2020unicoder,su2019vl,chenuniter,li2020oscar,tan2019lxmert,yu2020ernie,huang2020pixel} mainly take a two-step training pipeline, which consists of extracting semantic visual features by specific object detector and training the cross-modal pre-training model to align text and visual features. There are typically two lines of research about this topic. The first line uses a single-stream transformer architecture to model both image and text representations in a unified semantic space such as VLBERT~\citep{su2019vl}, UNITER~\citep{chenuniter} and OSCAR~\citep{li2020oscar}. In contrast, the other line adopts a two-stream Transformer architecture that first encodes the image and text modalities separately, and then fuses the cross-modal representations with another Transformer network, such as LXMERT~\citep{tan2019lxmert} and ERNIE-ViL~\citep{yu2020ernie}. Recently, VinVL~\citep{zhang2021vinvl} 
pre-trained a large-scale object-attribute detection model with much larger amounts of data on four public object detection datasets for further improving the performance.  In addition to image-text pairs, UNIMO~\citep{li2020unimo} also employed large scale of free text corpus and image collections for enhancing the cross-modal learning. These methods rely heavily on a task-specific bounding box (or region) based object detector, which impose unnecessary constraints on model designs and limit potential applications of existing vision and language systems. In this paper, Grid-VLP explores a new way for vision-language pre-training by utilizing the grid-based convolutional features, which skips the expensive region-related steps and is not restricted to the necessity of object detector. 


\section{Grid-VLP Approach}

\subsection{Model Architecture}
The architecture of Grid-VLP is shown in Figure~\ref{fig:framework}. Given a pair of aligned image and caption text, we first utilize a pre-trained CNN encoder to extract the grid features for the image, then a Transformer-based model is used to conduct cross-modal fusion directly between the token embeddings and image feature map. Different V+L pre-training tasks are designed to further enhance the cross-modal learning. To make up for the difficulty of modeling long grid feature sequence and accelerate training, we adopt a simple random grid sampling mechanism to select part of image feature map during cross-modal fusion. The architecture is flexible and no expensive region-related steps are involved. 

\subsection{Input Representations}
The input to Grid-VLP is an image and its related text (e.g. caption text). We first introduce the way to represent the text sequence and image. 

\paragraph{Sentence Embeddings} 
Similar to BERT~\citep{devlin2018bert}, each sentence is first split into a sequence of sub-words $\{w_{1},...,w_{m}\}$ by WordPiece tokenizer. Then each token $w_i$ is assigned three kinds of learnable embeddings: token, modal type and position embeddings. The three embeddings are summed and layer-normalized to represent input sentence as a sequence of embedding vectors $E_{emb}=\{e_{CLS},e_{1},...,e_{m},e_{SEP}\}$, where $[CLS]$ and $[SEP]$ are special tokens in BERT.

\paragraph{Image Grid Features} 
We use the traditional grid-based convolutional features instead of region-based features for representing an image. 
Starting from the initial image $v_{img}\in R^{3\times H_{0}\times W_{0}}$ with 3 color channels, a CNN-based image encoder generates a lower-resolution activation map $f_{img}\in R^{C\times H\times W}$ using the typical values as in Faster R-CNN~\citep{ren2015faster}: $C=2048$ and $H=\frac{H_0}{32}, W=\frac{W_0}{32}$. To make the CNN encoder aware of fine-grained semantics, we add a $1\times1$ RoIPool layer on top of the CNN encoder followed by two fully-connected layers and pre-train the encoder on Visual Genome dataset as in~\citep{jiang2020defense}. After pre-training, the CNN encoder is fixed as a grid feature extractor.  As the cross-modal fusion network expects a sequence as input, we collapse the spatial dimensions of $f_{img}$ into one dimension, resulting in a $HW\times C$ feature map $s_{img}$. Then we take a linear projection layer to reduce the channel dimension of the high-level feature map $s_{img}$ from $C$ to a smaller dimension $d$ so as to match the dimension of token embeddings. To distinguish between different modalities, we supplement the grid feature map with a learnable modal type embedding that are added to the output of linear projection layer. Finally, the sequential image representation $I_{img}=\{g_1, ..., g_{HW}\}$ can be seen as a $HW$ length of $d$-dimensional vector.

\subsection{Cross-modal Fusion}
Given the embeddings of the tokens for the sentence $\{e_i\}_{i=1}^m$ and the sequential image representations $\{g_j\}_{j=1}^n$, we adopt the Transformer encoder to learn cross-modal fusion between image grids and language tokens. To allow a fine-grained feature-level fusion, we concatenate the derived image features and text embeddings to construct the input sequence, which is formulated as: $\{e_{CLS},e_{1},...,e_{m},e_{SEP},g_1,...,g_{HW}\}$. To facilitate cross-modal understanding, we follow LXMERT~\citep{tan2019lxmert} and conduct three popular pre-training tasks\footnote{\small{Different from LXMERT, we do not have any vision pre-training task without the concept of region.}}, including Masked Language Modeling (MLM), Image-Text Matching (ITM) and Image Question Answering (QA). 
\paragraph{Random Grid Sampling} 
During pre-training, to further increase the difficulty of cross-modal pre-training tasks, we adopt a random sampling strategy as in~\citep{huang2020pixel} by randomly sampling a fixed number of grids for each image. At each iteration step, given the sequence of extracted grid features, we will dynamically sample a part from them and feed it into Transformer. In this way, we encourage the model to learn cross-modal relation with incomplete visual input, so as to enhance the robustness. Besides, it can largely accelerate the training process by reducing the total input sequence length. During fine-tuning, we still use the completed grid features to keep all the extracted visual information. 

\section{Experiments}

\begin{table*}
\centering
\small
\begin{tabular}{cl|c|ll|ll|ll}
\toprule
\multicolumn{2}{c|}{\multirow{2}{*}{Models}}      &
\multirow{2}{*}{Params} &
\multicolumn{2}{c|}{VQA} & \multicolumn{2}{c|}{NLVR2} & \multicolumn{2}{c}{GQA}  \\
\multicolumn{2}{l|}{}    &    & Test-dev & Test-std    &  Dev       &Test-P  &   Test-dev   & Test-std             \\
\midrule
\multirow{5}{*}{Single-stream}  & VLBERT & 110M   & 71.16    & -        & - & -            & - & -              \\
                              & UNITER & 110M     & 72.70    & 72.91        & 77.14 & 77.87             & - & -             \\
                              & OSCAR & 110M       & 73.16    & 73.61        & 78.07     & 78.36     & 61.19               & 61.23                         \\
                              & UNIMO & 110M       & 73.79    & 74.02        & -     & -     & -               & -                         \\
			     & VinVL   & 110M       & 75.95    & 76.12        & \textbf{82.05}     & \textbf{83.08}  & \textbf{65.05} & \textbf{64.65}             \\
\midrule
\multirow{4}{*}{Two-stream}    & ViLBERT & 221M & 70.55    & 70.92        & 67.40 & 67.00             & -          & -                 \\
                              & 12-in-1 & 221M & 73.15    & -        & - & - & -            & 60.65                     \\
                              & LXMERT & 183M          & 72.42    & 72.54        & 74.90     & 74.50     & 60.00                        & 60.33                \\
                              & ERNIE-ViL & ~210M  & 72.62    & 72.85        & - & - & -          & -           \\
\midrule
End2End & PixelBERT & 170M &74.45 &74.55 & 76.50 & 77.20 &- &- \\
\midrule
Our  Model             & Grid-VLP & 110M &  \textbf{76.05}   &   \textbf{76.18}    & 80.03       &     81.11  &       63.50      &       63.81          \\
\bottomrule
\end{tabular}
\caption{Evaluation Results on VQA, NLVR2 and GQA.}
\label{table:overall1}
\end{table*}

\subsection{Pre-training Dataset}
The same in-domain data as in LXMERT~\citep{tan2019lxmert} is used for pre-training. It consists of the image caption data from MS-COCO~\citep{lin2014microsoft}, Visual Genome~\citep{krishna2017visual}, and image question answering data from VQA v2.0~\citep{antol2015vqa}, GQA balanced version~\citep{hudson2019gqa} and VG-QA~\citep{zhu2016visual7w}. The total amount of the dataset is 9.18M image-and-sentence pairs on 180K distinct images.

\subsection{Implementation Details}
The maximum sequence length for the sentence is set as 20. The visual encoder is selected as ResNeXt with different sizes~\citep{xie2017aggregated} as in~\citep{jiang2020defense,huang2020pixel}. We resize the shorter side of the input image to 600, and limit the longer side to at most 1000. We select a fixed number of 100 random grids each time during pre-training~\footnote{\small{We also test with different number of selected grids such as 64 and 128, it does not make much difference.}}. We pre-train Grid-VLP with 12 layers of Transformer encoder. The basic settings of the Transformer is the same as BERT~\citep{devlin2018bert}. We pre-train Grid-VLP with a total batch size of 512 for 30 epochs on 8 V100 GPUs. We use the AdamW optimizor and set the initial learning rate as $10^{-4}$. 


\subsection{Main Results on Downstream Tasks}
We compare Grid-VLP model against all the state-of-the-art VLP models of the comparable model size on three popular V+L understanding tasks: VQA~\citep{antol2015vqa}, NLVR2~\citep{suhr2018corpus} and GQA~\citep{hudson2019gqa}. The detailed task descriptions can be referred in LXMERT~\citep{tan2019lxmert}. The results are shown in Table~\ref{table:overall1}. We can see that with only in-domain pre-training data on MS-COCO and Visual Genome, Grid-VLP can outperform almost all the other existing VLP methods~\footnote{\small{Only one exception is VinVL, which uses large amounts of objection detection data, out-of-domain pre-training data and fine-tuning tricks such as using unbalanced data on GQA.}} by a large margin, without using any complicated strategies such as data augmentation, adversarial training and knowledge enrichment. Grid-VLP even outperforms an end-to-end method PixelBERT, which directly optimizes on pixel level. It shows the effectiveness of the proposed approach for conducting VLP on a grid feature level, which provides a brand new way for vision-language pre-training.


\subsection{Influence of Image Encoder}
\begin{table}
\centering
\small
\begin{tabular}{l|c|c|c} 
\toprule
Backbone & Avg Time (ms) &VQA  & NLVR2 \\
\midrule
x50  & 106 & 74.34& 78.01\\
x101 & 153 & 75.18& 78.95\\
x152 & 221 & \textbf{76.05}& \textbf{80.03}\\
\midrule
x152 (region) & 572 & 75.12& 79.14\\
\bottomrule
\end{tabular}
\caption{Results of different image encoder architectures on VQA and NLVR2 development set.}
\label{table:size} 
\end{table}

As the image encoder is critical for extracting grid features, we further study the importance of the image encoder by changing different ResNet visual backbone layers. To allow a fair comparison between grid features and region features, we also add one experiment that using Faster R-CNN~\citep{ren2015faster} with the the same ResNeXt backbone to extract region features of 100 objects. The result is shown in Table~\ref{table:size}. We can see that using more complicated visual backbones can contribute to the final performance gain, which proves the importance of the image encoder for cross-modal understanding. Besides, given the same settings with only the difference of extracted feature type, the performance of grid features can even outperform region-based features with a much shorter feature extraction cost. 

\subsection{Impact of Input Image Size}
\begin{table}
\centering
\small
\begin{tabular}{cc|c|c|c} 
\toprule
 \multicolumn{2}{c|}{Input Size} & \multirow{2}{*}{Speedup} & \multirow{2}{*}{VQA} & \multirow{2}{*}{NLVR2}\\
shorter side & longer side & & & \\
\midrule
448& 448 & 3x & 74.71 & 79.22\\
448 & 746 & 2x & 75.64 & 79.72 \\
600 & 1000 & - & \textbf{76.05} & \textbf{80.03} \\
\bottomrule
\end{tabular}
\caption{Impact of input image size on VQA and NLVR2 development set.}
\label{table:img_size} 
\end{table}

As the sequence length of the grid features is determined by the image size $H$ and $W$. Therefore, the final sequence length of the input to the transformer also largely depends on the image size, which is important to control effectiveness and efficiency tradeoff. We further analyze the impact of input image size to Grid-VLP as shown in Table~\ref{table:img_size}. From the results, we can see that Grid-VLP benefits from larger image sizes as input, where more fine-grained information can be extracted. Moreover, resizing the image to a smaller size can significantly improve the inference speed without decreasing performance much, e.g., no more than 1\% for 2 times speedup. 


\section{Conclusion}
In this paper, we revisit grid features as an alternative to the dominant bottom-up region features for vision-language pre-training. We propose a rather simple yet effective grid-based VLP method, and find that with larger and well pre-trained visual backbones, our Grid-VLP method can also outperform most competitive region-based VLP methods by equipped with the grid-based convolutional features. We hope our findings can help further advance the progress of vision-language pre-training and potentially provide new perspectives to vision-language pre-training.  

\bibliography{anthology,custom}
\bibliographystyle{acl_natbib}




\end{document}